\documentclass[prb,twocolumn,showpacs]{revtex4}
\usepackage{amssymb}
\usepackage{graphicx}

\begin{document}

\title{Spin Resonance and dc Current Generation in a Quantum Wire}
\author{V. L. Pokrovsky}
\affiliation{Department of Physics, TexasA\&M University and Landau Institute for
Theoretical Physics}
\author{W. M. Saslow}
\affiliation{Department of Physics, TexasA\&M University}

\begin{abstract}
We show that in a quantum wire the spin-orbit interaction leads to a narrow
spin resonance at low temperatures, even in the absence of an external
magnetic field. Resonance absorption by linearly polarized radiation gives a
dc spin current; resonance absorption by circularly polarized radiation
gives a dc electric current or magnetization. 
\end{abstract}

\maketitle



\textbf{Introduction. }ESR in 3d metals subject to an external magnetic
field is associated with spin-flip processes of electrons with momenta
between the up-spin and down-spin Fermi-spheres. 
The spin-flip energy (Zeeman energy) is the same for any electron because
the ac-field is almost uniform. For a sufficiently strong spin-orbit (SO)
interaction this ESR resonance is smeared out, since spin then couples to
the lattice and the Zeeman energy is no longer uniquely determined. Shechter 
{\textit{e}t al} \cite{finkel} noted that in a thin film the SO interaction
in the Rashba form \cite{rashba} leads to what they called a \textit{chiral
resonance.} There the stationary states are characterized by chirality, i.e.
the direction of spin parallel or antiparallel to the vector $\hat{z}\times 
\mathbf{p}$, where $\hat{z}$ is the normal to the film and $\mathbf{p}$ is
the momentum. Spin-flip from one chirality to another requires energy $%
E_{sf}=2\alpha p$, where $\alpha $ is the Rashba SO constant. Since in the
ground state this process is possible only for electrons whose momenta are
located between two Fermi-circles corresponding to different chiralities,
the spin-flip energy is distributed in a narrow band centered at $%
E_{sf}=2\alpha p_{F}$ with bandwidth $\Delta =4m\alpha ^{2}$. Unfortunately,
the Dresselhaus SO interaction \cite{dressel} strongly broadens the
resonance. The Dresselhaus interaction in 2d can be written as $H_{D}=\beta
\left( p_{x}\sigma _{x}-p_{y}\sigma _{y}\right) $, whereas the Rashba
interaction is $H_{R}=\alpha \left( p_{x}\sigma _{y}-p_{y}\sigma _{x}\right) 
$. Together they give a spin-dependent contribution to the Hamiltonian with
effective magnetic field whose magnitude depends not only on the momentum
magnitude $p$ but also on its azimuthal angle $\varphi $, with spin-flip
energy $E_{sf}=2p\sqrt{\alpha ^{2}+\beta ^{2}+2\alpha \beta \sin 2\varphi }$%
. This leads to a bandwidth $\Delta =2p_{F}\min \left( \alpha ,\beta \right) 
$ of the same order of magnitude as the spin-flip energy. In Ref.~%
\onlinecite{finkel} the authors proposed to avoid this broadening by
choosing a specific growth direction or by decreasing the electron density.
Both proposals require a sophisticated experimental technique.

This work builds on the fact that anisotropic broadening vanishes for a
quantum wire where, as we will show, the direction of the effective magnetic
field in the wire is the same for all momenta. In a quantum wire with only
one or a few channels new effects that are absent in 2 and 3 dimensions are
predicted: (1) a circularly polarized magnetic field produces a finite
magnetization and a (weak) spin-polarized electric current; (2) a linearly
polarized ac magnetic field produces either a constant \textit{spin current}
or separation of opposite magnetic moments in the wire. Independent of
polarization of the ac field, a \textit{resonant heating} (RH) takes place.
The (resonant) spin-polarized electric current is reminiscent of the
(non-resonant) photo-galvanic effect;\cite{ivchenko-pikus,belinicher} both
appear space inversion symmetry is violated normal to the wire. The
resonance frequency and the magnitude of the effect can be controlled easily
with the external magnetic field and gate voltage $V_{g}$. These two effects
are experimentally observable. The wire can be formed from a semiconducting
film or heterojunction by a proper configuration of the gate electrodes \cite%
{1d-from-2d} or by the growth process.\cite{nanowires} In the latter case
the wire must be deposited on a substrate that violates reflection symmetry.
The 1d geometry, together with quantization of the transverse motion,
strongly suppresses the most effective Dyakonov-Perel mechanism of spin
relaxation,\cite{dyakonov-perel} thus stabilizing the resulting
non-equilibrium state.

\textbf{Electronic spectrum and eigenstates.} Consider the 
type III-V semiconductors GaAs and InGaAs, and only their electron bands, to
avoid complications associated with degeneracy of the hole band. The
electron density is assumed to be sufficiently large and the temperature
sufficiently low to ensure 
a degenerate Fermi gas. We neglect the electron-electron interaction and
assume that the wire is narrow enough to exclude multiple channels. 
Interactions, i.e. Luttinger liquid effects in a 1d electron system with a
strong Coulomb interaction, \cite{zulicke} will be neglected. 
Let the $x$-axis be directed along the wire. With $\left\langle
p_{y}\right\rangle =0$ due to quantization, the total Hamiltonian, including
both Rashba and Dresselhaus interactions, reads:%
\begin{equation}
H=\frac{p^{2}}{2m}+\left( \alpha \sigma _{x}+\beta \sigma _{y}\right) p,
\label{Hamiltomian}
\end{equation}%
where $m$ is the effective mass, $p=p_{x}$ is the only non-trivial momentum
component, and the $\sigma _{x,y}$ are the usual Pauli spin matrices. The
eigenvalues of Eq.~(\ref{Hamiltomian}) are:%
\begin{equation}
E\left( p,\sigma \right) =\frac{p^{2}}{2m}+\sigma \gamma p,~\gamma =\sqrt{%
\alpha ^{2}+\beta ^{2}}  \label{eigenvalues}
\end{equation}%
where $\sigma =\pm 1$ are the eigenvalues of the operator $\sigma _{\tilde{x}%
}=\frac{\alpha \sigma _{x}+\beta \sigma _{y}}{\sqrt{\alpha ^{2}+\beta ^{2}}}$%
. The spin quantization axis $\tilde{x}$ is tilted at an angle $\phi
=\arctan \frac{\beta }{\alpha }$ to the wire.\cite{1d-v} The eigen-spinors
corresponding to the state $(p,\sigma)$ are:%
\begin{equation}
\chi _{p\sigma }=\frac{1}{\sqrt{2}}\left( 
\begin{array}{c}
\sigma e^{-i\phi /2} \\ 
e^{i\phi /2}%
\end{array}.%
\right)  \label{eigenspinors}
\end{equation}%
\noindent\ 

In the ground state particles with chirality $\sigma $ fill a momentum
interval from $p_{\sigma -}$ to $p_{\sigma +}$. Here 
\begin{equation}
p_{\sigma \tau }=\tau p_{F}-\sigma m\gamma, \quad p_{F}=mv_{F}= \frac{\pi
\hslash n_{1}}{2},  \label{p-sigma}
\end{equation}%
where $\tau =\pm 1$ distinguishes right (R) and left (L) movers, 
and $n_{1}$ is the 1d electron density. All states in the interval $\left(
p_{--},p_{++}\right) $ are occupied twice; the states in the intervals $%
\left( p_{++},p_{-+}\right) $ and $\left( p_{+-},p_{--}\right) $ are singly
occupied (Fig. 1).\ \emph{\ } A net spin-flip is possible only in the singly
occupied intervals, and requires energy $E_{sf}=2\gamma \left\vert
p\right\vert $. The \textquotedblleft length\textquotedblright\ of each
singly-occupied interval is $2m\gamma $. Assuming that $m\gamma \ll \hslash
n_{1}$, we find that the spin-flip energies are confined to a narrow band of
width $\Delta =4m\gamma ^{2}$ centered at $E_{sf}^{\left( 0\right) }=2\gamma
p_{F}$. Spin-flip processes can be excited by an external field with
resonance frequency and width 
\begin{equation}
\omega_{r}=2\gamma p_{F}/\hslash =\pi \gamma n_{1}, \quad \Delta
\omega_{r}=4m\gamma ^{2}/\hslash.  \label{res-width}
\end{equation}
The temperature $T$ must satisfy $T<\hbar\omega_{r}/k_{B}$ 
to avoid thermal smearing.%

\begin{figure}[htbp]
\centering
\includegraphics[width=3.2in]{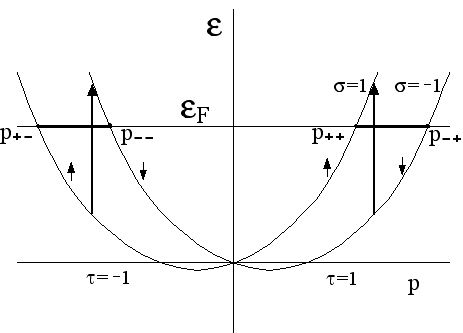} 
\caption{Energy \textit{vs} momentum in one dimension, for spin up ($\protect%
\sigma=1$) and spin down ($\protect\sigma=-1$) bands. Small arrows indicate
spin direction. Thick horizontal bars indicate states in the lower band that
can be excited to the upper band. Two allowed transitions are indicated by
long vertical arrows.}
\label{fig: RashbaBands}
\end{figure}

\textbf{Linearly polarized ac field. }Let an ac magnetic field $B\left(
t\right) =B_{0}\left( t\right) e^{-i\omega _{0}t}+B_{0}^{\ast }\left(
t\right) e^{i\omega _{0}t}$ be linearly polarized along $z$, with spectral
intensity $I\left( \omega \right) $ distributed in a narrow {frequency
interval} $\Delta \omega $ near the central frequency $\omega _{0}$. The
envelope amplitude $B_{0}\left( t\right) $ describes a stationary random
process with correlation time $\tau _{B}=\frac{2\pi }{\Delta \omega }$.
Averaged over {a} time interval $t^{\prime}$ satisfying $2\pi/\omega_{r}\ll
t^{\prime}\ll \tau_{B}$, 
the two time correlation function satisfies 
$\overline{B_{0}^{\ast }\left( t\right) B_{0}\left( t^{\prime }\right) }%
=\left( 2\pi \right) ^{-1}\int_{-\infty }^{\infty }I_{\omega }e^{i\omega
\left( t-t^{\prime }\right) }d\omega $. 
As {interaction Hamiltonian} we take $H_{B}=-g\mu _{B}B\left( t\right)
\sigma _{z}$.  With $\left( \chi _{p+},\sigma _{z}\chi
_{p-}\right) =1$ for the two states $\chi _{p\sigma }$ given by Eq. (\ref%
{eigenspinors}), lowest order time-dependent
perturbation theory gives the {spin-flip transition rate} for an electron with
momentum $p$ as
\begin{equation}
w=\frac{g^{2}\mu _{B}^{2}}{\hslash ^{2}}I_{(2\gamma \left\vert p\right\vert
/\hslash) -\omega _{0}}.  \label{rate}
\end{equation}%
In order of magnitude 
$I_{\omega }\approx \pi \overline{B^{2}}/\left( \Delta \omega \right) $, so 
\begin{equation}
w\approx\frac{\pi g^{2}\mu _{B}^{2}\overline{B^{2}}}{\hslash^{2}\Delta \omega%
}.  \label{w}
\end{equation}


{Spin-flip pumping by a linearly polarized ac} magnetic field creates
non-equilibrium right-moving {excitations} with spin up and left-moving {%
excitations} with spin down. Simultaneously it creates holes in the {%
Fermi-sea} with the same momenta as {the} excitations. Their 1d densities $%
n_{ex}$ are determined by steady-state {balance of the} spin pumping rate $w$
of (\ref{w}) \textit{vs} the spin relaxation time $\tau_{sf}$. 
With $n_{1}\gamma/v_{F}$ the saturation density of right-moving up spins
from the down-only momentum states, we have 
\begin{equation}
n_{ex}\approx\frac{n_{1}\gamma }{v_{F}}\min \left( w\tau _{sf},1\right).
\label{n_ex}
\end{equation}
The equal number of left-moving down spins cause the electric current and
magnetization to be zero, but the spin current is non-zero.

In the absence of back-scattering the spin current is $I_{s}=2n_{ex}v_{F}$,
in units of $g\mu_{B}$. However, back-scattering without spin flip {reverses
the velocity and tends to destroy the} spin current. Let $\tau_{b}$ be the
back-scattering time without spin-flip, determined by collisions with
non-magnetic impurities. Since $\tau_{b}\ll (\tau_{sf}, w^{-1})$, when $%
n_{ex}$ is saturated $I_{s}$ is suppressed by the factor $w\tau_{b}$, and
when $n_{ex}$ is not saturated $I_{s}$ is suppressed by the factor $%
\tau_{b}/\tau_{sf}$. These two cases can be written as a single equation: 
\begin{equation}
I_{s}=2n_{ex}v_{F}\frac{w\tau _{b}}{\min \left( w\tau _{sf},1\right) }%
=2n_{1}\gamma w\tau _{b}.  \label{spin-curr}
\end{equation}%
%
%

In the ballistic regime $\tau _{b}$ must be replaced by the {time of flight}%
. If the spin current does not completely penetrate into {the} connecting
leads, {then} the accumulation of positive spins at the right end, and {of}
negative spins at the left end of the wire, can be observed using {the}
magnetic Kerr effect, as {was} done {for the} spin-Hall effect.\cite%
{awshalom} 
For a wire of length $L$ and magnetization $\pm M$ at the ends we take $%
I_{s}\sim D M/L$, with spin diffusion constant $D\sim v_{F}^{2}\tau_{b}$.
Use of \ref{spin-curr} then yields a magnetization per particle, in units of 
$g\mu _{B}${,} of $\pm \frac{\gamma wL}{v_{F}^{2}}$. It is rather small.

\textbf{Circularly polarized ac field. }If the ac magnetic field is
circularly polarized, it produces spin-flips only for spins of one direction{%
; for example, spin} down. In this case an excess of right-moving spin up
electrons appears. Therefore, the circularly polarized magnetic field
produces a finite magnetic moment per unit length 
\begin{equation}
M_{1}=2g\mu _{B}n_{ex}=2g\mu _{B}n_{1}\frac{\gamma }{v_{F}}\min \left( w\tau
_{sf},1\right)  \label{magnetization}
\end{equation}%
The magnetization is stable with respect to {back scattering} since these
processes {rarely} reverse spins, but to observe it {the} experiment must
ensure that spins do not flow {through} the ends of the wire. The same ac
field generates {both} dc spin and electric currents. {Since now there is
only one circular component of ac magnetic field, the 
spin current} 
is half that given by Eq. (\ref{spin-curr}).

Now consider the electric current. In {equilibrium} the current from
particle $(p,\uparrow)$ 
cancels the current from particle $(-p,\uparrow)$, since they have opposite
velocities (see Fig. 1). When particle $(p,\uparrow)$ flips its spin, it
simultaneously changes its velocity by $2\gamma $. Therefore, the total
electric current is {no longer} zero. A balance argument yields:%
\begin{equation}
I_{e}=2en_{ex}\gamma \frac{w\tau _{b}}{\min \left( w\tau _{sf},1\right) }%
=2en_{1}\frac{\gamma ^{2}}{v_{F}}w\tau _{b}.  \label{el-curr}
\end{equation}%
For an isolated wire, with zero net current flow, a voltage bias $V=-I_{e}R $
develops across the wire.

\textbf{Resonant heating (RH). }{For} each spin-flip {an} electron absorbs
energy $2\gamma p_{F}$, so the power absorbed is $W=2\gamma p_{F}w$ per
excited particle. The average energy absorbed by a particle during the
spin-relaxation time is about $E_{abs}=2\gamma p_{F}\min \left( w\tau
_{sf},1\right) $, so the energy absorbed {per} unit length {equals} $%
Q=E_{abs}n_{1}[\min \left( \hslash \Delta \omega ,2\gamma
p_{F}\right)/v_{F}p_{F}]$. (The bracketed factor is the fraction of
electrons {that} absorb the energy.) The absorbed energy turns into {heat}
since usually {spin} relaxation proceeds {more slowly} than {energy}
relaxation. {We call this} resonance heating (RH). The temperature change {is%
} $\Delta T=Q/C$, where $C\approx n_{1}T/\varepsilon _{F}$ is the specific
heat per unit length of {the} 1d degenerate Fermi gas. This relation and (%
\ref{w}) 
yield 
\begin{equation}
\Delta T\approx\frac{2\pi \varepsilon _{F}}{T}\mathrm{min}(w\tau_{sf},1).
\label{heating1}
\end{equation}
{The RH can be determined from the temperature-dependence of the resistance.}

\textbf{Numerical estimates}{. }All numerical estimates are made for In$%
_{0.53}$Ga$_{0.47}$As; we take $m=4.3\times 10^{-29}$~g; $\alpha =1.08\times
10^{6}$~cm/s, and $g=-0.5$.\cite{dorokhin,note} 
For a typical value of $\hslash \Delta \omega _{r}$, (\ref{w}) yields, for $B
$ in G, $w=0.42\times 10^{4}B^{2}$~s$^{-1}$. The value {of} $B^{2}$
accessible {at terahertz frequencies} is $\sim $~30G$^{2}$ and may be
increased.\cite{highfield} Thus, it is feasible to reach $w\sim (4\times
10^{3}-4\times 10^{5})$~s$^{-1}$. Below we show that the spin flip time $%
\tau _{sf}$ satisfies $w\tau _{sf}\gtrsim 1$, 
so the probability of spin excitation for electrons involved in the
resonance is of the order of 1. A typical 2d electron density is $%
n_{2}\approx 2\times 10^{12}$/cm$^{2}$. We take a wire of width $a=10$~nm,
so $n_{1}=n_{2}a=2\times 10^{6}$/cm. 
The mean free path for the back scattering is $l_{b}=\left( n_{i}R\right)
^{-1}$, where $n_{i}$ is the one-dimensional density of impurities and $R$
is the reflection coefficient. We assume that $n_{i}$ coincides with $n_{1}$%
; all impurities are positively charged and cause a weak reflection, which
we estimated in the Born approximation, assuming the dielectric constant to
be about 10. Our estimate for reflection coefficient is $R\sim 10^{-3}\div
10^{-2}$. Thus, $l_{b}=10^{2}-10^{3}/n_{1}\sim 10^{-3}-10^{-4}$~cm. so
electrons in a shorter wire are in the ballistic regime. Further, $\tau
_{b}=l_{b}/v_{F}=10^{-10}-10^{-11}$~s. By (\ref{el-curr}) $I_{e}$ is 1-10~pA
and thus is observable. 

\textbf{Spin relaxation}. {Relaxation processes play an} extremely important
role in spin {resonance} phenomena. 
The most important spin relaxation mechanism in {2d and 3d} semiconductors
is {Dyakonov-Perel} impurity scattering.\cite{dyakonov-perel} It {is
effective} in the diffusive regime $\omega _{s}\tau _{i}\ll 1$, where $%
\omega _{s}$ is the characteristic spin precession frequency 
and $\tau _{i}$ is the {electron-impurity collision time.} 
Here $\omega _{s}=\omega_{r}=2\gamma p_{F}/\hslash $. {Spin relaxation
accompanying} elastic collisions with impurities is associated with {spin
rotation due to the effective magnetic field acting during the time between
collisions.} Since in {1d} the direction of the effective magnetic field is
independent of {the} momentum, the spin has only two distinct orthogonal
eigenstates. {Thus, for a quantum wire a} collision with a non-magnetic
impurity {cannot flip spin}. (For $d>1$ the {spin eigenstates} depend on the
momentum direction. Therefore the spin states before and after collision are
not orthogonal.) 
Spin-flip due to collisions with magnetic impurities is relatively unlikely
since the concentration of magnetic impurities is low. The corresponding
spin-flip, or spin-relaxation, time is $\tau _{sf}^{\left( mi\right)
}=\left( v_{F}n_{mi1}P_{mi}\right) ^{-1}$, where $n_{mi1}$ is the 1d
concentration of magnetic impurities and {$P_{mi}$} is the probability of {%
spin} flip at a collision. For realistic magnetic impurity concentrations $%
\tau _{sf}^{\left( mi\right) }$ is rather long (about 3~ms); other sources
of relaxation produce {a} much shorter $\tau _{sf}$.

{Spin-flip also} may be associated with emission or absorption of a phonon.
We have considered $\tau _{sf}^{\left( ph\right) }$ for: (i) a wire with no
acoustic contact to the substrate (a good approximation for {an individually}
grown wire or a nanotube); (ii) {a wire with a strong acoustic coupling} to
the substrate (as for a wire formed by special gate electrodes in {the} 2d
electron layer). In the first case the phonons are also one-dimensional; in
the second case the phonons propagate in 3d space and have {many} more
degrees of freedom. We estimate that 
$\tau _{sf}^{\left( ph,1d\right) }\sim 2$~ms for 1d phonons and $%
\tau_{sf}^{\left( ph,3d\right) }\sim 10^{-8}$~s for 3d phonons.\cite%
{EnergyRelax}

\textbf{Energy and momentum relaxation}. Transitions without spin flip
produce {energy} and momentum relaxation. First consider the decay of an
electron (particle) excitation. The simplest {decay process,} into two
particles and one hole{,} is forbidden in {1d} by {energy and momentum
conservation; the simplest allowed process is the decay of a right-moving
particle R into three particles and two holes, with one particle and one
hole each left-moving (L)}.\cite{khodas} {For the} Coulomb interaction the {%
inverse decay} time is proportional to the fourth power of the deviation of {%
the} initial particle momentum from {the} corresponding Fermi boundary $%
p_{++}$. {A} numerical estimate for the relaxation time associated with the
decay gives $\tau _{d}\sim 2.5\times 10^{-4}$~s.\cite{EnergyRelax} We {also}
calculated the rate of transition due to emission-absorption of phonons in
an acoustically isolated wire. With $U$ the deformation potential, $u$ the
sound velocity, and $\mathcal{A}$ the wire's cross-section, we find\cite%
{EnergyRelax} 
\begin{equation}
\Gamma _{s}=\frac{2ma^{3}U^{2}}{\hslash ^{2}\mathcal{A}Mu},  \label{Gamma}
\end{equation}
$U=16$~eV, $u=3.3\times 10^{5}$~cm/s,\cite{sugaya} and $\mathcal{A}=100$~nm$%
^{2} $, $\Gamma _{s}=1.19\times 10^{11}$~s$^{-1}$, gives an energy
relaxation time $\tau _{s}\sim 10^{-11}$~s. This {rapid} relaxation ensures
the transformation of the absorbed energy into {heat} and {prevents}
possible Rabi oscillations. {When} the wire is acoustically coupled {to} the
substrate{,} the relaxation proceeds even faster $(\tau _{s}\sim
10^{-14}-10^{-12}$~s). However, in this case the absorbed energy does not
heat {the electrons, but rather the substrate lattice, producing a barely
measurable effect.} The details of {the theory of these relaxation processes
will be published elsewhere.}\cite{EnergyRelax}


\textbf{Influence of constant magnetic field. }{An external constant
magnetic field $\mathbf{B}_{0}$} changes the effective magnetic field. As a
result the resonance frequency {becomes} 
\begin{equation}
\omega _{r}=\sqrt{\left( \omega _{r0}+\frac{g\mu _{B}B_{0\Vert }}{\hslash }%
\right) ^{2}+\left( \frac{g\mu _{B}B_{0\bot }}{\hslash }\right) ^{2}},
\label{omega'}
\end{equation}
where $\omega_{r0}=2\gamma p/\hslash $ is the resonance frequency in the
absence of {$\mathbf{B}_{0}$, with components $B_{0\Vert }$ and $B_{0\bot }$}
parallel and perpendicular to the effective spin-orbit magnetic field. By
proper choice of $B_{0\Vert }$ it is possible to decrease $\omega_{r}$
significantly. However, this also reduces the difference $p_{-+}-p_{++}${,}
and thus the temperature below which the resonance is not too broadened.

\textbf{Influence of gate voltage. }The effective Rashba interaction $\gamma
=\sqrt{\alpha ^{2}+\beta ^{2}}$ varies linearly with the gate voltage $V_{g}$%
, so that $\gamma$ can be varied. In addition, the Fermi-momentum and the
gyromagnetic ratio\emph{\ }{also vary linearly with $V_{g}$.} Thus, the
resonance frequency \emph{is a }quadratic polynomial in the $V_{g}${.
However,} the relative width of {the} resonance is {independent of $V_{g}$}.
The spin and electric currents, as well as the absorbed heat, also vary as
cubic polynomials of $V_{g}$ ($g, v_{F}, n_{1}, \tau_{b}^{-1}$ all are
linear in $V_{g}$). Thus, by increasing $V_{g}$ it is possible, in
principle, to increase the currents.

\textbf{Conclusions}. {We have} predicted a new effect: spin resonance in a
quantum wire with no external magnetic field. It occurs because for a
degenerate Fermi gas in one dimension the direction of the effective
spin-orbit magnetic field is almost independent of momentum. The resonance
frequency is typically in {the} terahertz region. Its relative width {%
depends linearly on the Dresselhaus and Rashba} spin-orbit constants. We
predict that {on resonance a linearly polarized ac magnetic field produces}
a dc spin current, whereas {a} circularly polarized magnetic field produces {%
a} completely spin polarized {dc} electric current of the order of pA. If
the wire is not connected to an electric circuit, the circularly polarized
field produces a dc magnetization. Observation of these effects is
experimentally feasible. In all cases the resonance heating of the wire {may
be measureable}, but only if the wire is acoustically decoupled {from} the
substrate. The amplitude of these effects can be easily {controlled by the
external constant magnetic field and by the gate voltage, thus suggesting}
technological applications.

\textbf{Acknowledgements}. We {thank} A.M. Finkelstein and M. Khodas for {%
discussions and for pointing out their work.\cite{finkel}} {This work was
supported by the Department of Energy} under grant DE-FG02-06ER46278.


\begin{thebibliography}{99}
\bibitem{finkel} A. Shekhter, M. Khodas, and A. M. Finkelstein, Phys. Rev. B 
\textbf{71}, 165329 (2005).

\bibitem{rashba} E. I. Rashba, Fiz. Tverdogo Tela, \textbf{2}, 1224 (1960)
[Sov. Phys. Solid \textbf{2}, 1109 (1960)]; Yu. A. Bychkov and E. I. Rashba,
Pis'ma v ZhETF \textbf{39, }66 (1984); [JETP Lett. \textbf{39}, 78 (1984)].

\bibitem{dressel} G. Dresselhaus, Phys. Rev. \textbf{100}, 580 (1955).

\bibitem{ivchenko-pikus} E. I. Ivchenko and G. E. Pikus, Pis'ma v ZhETF 
\textbf{27, }640 (1978) [Sov. Phys. JETP Lett. \textbf{27, }604 (1978)].

\bibitem{belinicher} V. Belinicher, Phys. Lett. A \textbf{66}, 213 (1978).

\bibitem{1d-from-2d} T. Schaepers, J. Knobbe, A.V. Guzenko, Phys. Rev. B 
\textbf{69}, 235323 (2004). 

\bibitem{nanowires} S. B. Kim et al., J. Cryst. Growth \textbf{201-202},
1283 (1999); O. Zsebk et al., Nanotechnology \textbf{12}, 32 (2001)
Jong-Horng Dai et al., Nanotechnology \textbf{16}, 407 (2005); C. L. Zhang
et al, Nanotechnology \textbf{16,} 1379 (2005); M. A. Verhejen et al., J.
Am. Chem. Soc. \textbf{128}, 353 (2006).

\bibitem{dyakonov-perel} M. I. Dyakonov and V. I. Perel, ZhETF \textbf{60},
1954 (1971); [Sov. Phys. JETP \textbf{33, }1053 (1971); M. I. Dyakonov, 
\textit{Basics of Semiconductor and Spin Physics}, in \textit{Spin Physics
in Semiconductors, }ed. M. I.\ Dyakonov, Springer Series in Solid-State
Sciences, Vol. 157 (2008).

\bibitem{zulicke} W. Ha\"ussler, Phys. Rev. B 63, 121310(R) (2001); M.
Governale and U. Z\"{u}licke, Solid State Communications \textbf{131}, 581
(2004).

\bibitem{1d-v} $v\equiv\partial H/\partial p=p/m+\gamma\sigma_{\tilde{x}}$
is commutes wth both $p$ and $\sigma_{\tilde{x}}$. Thus an ac $E$-field
along the wire, which couples to $v$, cannot flip spins, in contrast to the
2d case. See Ref.\onlinecite{finkel}.

\bibitem{awshalom} Y. K. Kato, R. C. Myers, A. C. Gossard, D. D. Awschalom,
Science \textbf{306},1910 (2004). 

\bibitem{dorokhin} M. V. Dorokhin et al, J. Phys. D, Appl. Phys. \textbf{41,}
245110 (2008).

\bibitem{note} For simplicity we neglect the anisotropy of the g-factor in a
wire. Experimentally it is not small (R. Danneau et al, Phys. Rev. Lett. 
\textbf{97}, 026403 (2006)) and must be taken in account in a more accurate
theory.

\bibitem{highfield} M. Sherwin, Nature \textbf{240}, 131 (2002); A. Deminger
and A. S. Renner, Laser Focus World, Jan. 2008, p. 111; M. C. Hoffman et al,
arXiv 0904.2516; \textit{Terahertz sources and Systems}, R. Harrison and D.
Lippens, eds., Nato Science Series, Springer, 2000.


\bibitem{EnergyRelax} V. Pokrovsky and W. M. Saslow (unpublished).

\bibitem{khodas} M. Khodas, M. Pustilnik, A. Kamenev, and L.I. Glazman,
Phys. Rev. B \textbf{76}, 155402 (2007).

\bibitem{sugaya} T. Sugaya et al, Appl. Phys. Lett. \textbf{81}, 727 (2002).
\end{thebibliography}
\end{document}